\documentclass[%
 reprint,
 amsmath,amssymb,
 aps,
 prb,
]{revtex4-1}

\usepackage{dcolumn}% Align table columns on decimal point
\usepackage{bm}% bold math
    \usepackage{amsmath,amssymb, mathrsfs}
\usepackage[colorlinks=true,bookmarks=false,citecolor=blue,urlcolor=blue]{hyperref} %pdflatex

\usepackage{graphicx}
\usepackage{float}
\usepackage{xcolor}

\begin{document}

\title{Topological Dissipation in a Time-Multiplexed Photonic Resonator Network}

\author{Christian Leefmans$^{1,\ast,\dagger}$, Avik Dutt$^{2,\ast}$, James Williams$^3$, Luqi Yuan$^4$, Midya Parto$^3$, Franco Nori$^{5,6}$, Shanhui Fan$^2$, Alireza Marandi$^{1,3,\dagger}$\\
\textit{$^1$Department of Applied Physics, California Institute of Technology, Pasadena, CA 91125, USA.\\
$^2$Department of Electrical Engineering, Stanford University, Stanford, CA 94305, USA.\\
$^3$Department of Electrical Engineering, California Institute of Technology, Pasadena, CA 91125, USA.\\
$^4$State Key Laboratory of Advanced Optical Communication Systems and Networks, School of Physics and Astronomy, Shanghai Jiao Tong University, Shanghai 200240, China.\\
$^5$Theoretical Quantum Physics Laboratory, RIKEN, Wako-shi, Saitama 351-0198, Japan.\\
$^6$Department of Physics, University of Michigan, Ann Arbor, Michigan 48109-1040, USA.\\
$^\ast$These authors contributed equally to this work.}\\
$^\dagger$Email: \href{mailto:cleefman@caltech.edu}{cleefman@caltech.edu}, \href{mailto:marandi@caltech.edu}{marandi@caltech.edu}
}

\date{\today}

\maketitle

\textbf{Topological phases feature robust edge states that are protected against the effects of defects and disorder~\cite{hasan_colloquium_2010}. The robustness of these states presents opportunities in fields such as condensed matter, ultracold atoms~\cite{cooper_cold_atoms_2019}, phononics~\cite{liu_topological_phononics_2020}, and photonics~\cite{ozawa_topological_2019} to design technologies that are tolerant to fabrication errors and resilient to environmental fluctuations. While most topological phases rely on conservative, or Hermitian, couplings, recent theoretical efforts have combined conservative and dissipative couplings to propose new topological phases for ultracold atoms~\cite{diehl_topology_2011} and photonic amplifiers~\cite{wanjura_topological_2020}. However, the topological phases that arise due to purely dissipative couplings remain largely unexplored~\cite{bardyn_topology_dissipation_2013}. Here we realize dissipatively coupled vedeviations of two prominent topological models, the Su-Schrieffer-Heeger (SSH) model~\cite{su_solitons_1979} and the Harper-Hofstadter (HH) model~\cite{harper_1955,hofstadter_1976}, in the synthetic dimensions of a time-multiplexed photonic resonator network.  We observe the topological edge states of the SSH and HH models, measure the SSH model’s band structure, and induce a topological phase transition between the SSH model’s trivial and topological phases. In stark contrast with conservatively coupled topological phases, the topological phases of our network arise from bands of dissipation rates that possess nontrivial topological invariants (i.e. topological dissipation), and the edge states of these topological phases exhibit isolated dissipation rates that occur in the gaps between the bulk dissipation bands. Our results showcase the ability of dissipative couplings to break time-reversal symmetry: a phenomenon that enables us to realize nonzero Chern numbers with an effective magnetic field. We expect that our demonstration of robust topological edge states with isolated dissipation rates may inspire new designs for open quantum systems and photonic devices such as mode locked laser and optical computing architectures. Moreover, our time-multiplexed network, with its ability to implement multiple synthetic dimensions, dynamic and inhomogeneous couplings, and time-reversal symmetry breaking synthetic gauge fields, offers a flexible and scalable architecture for future work in synthetic dimensions~\cite{ozawa_topological_synthetic_2019}.}

\begin{figure*}
    \centering
    \includegraphics[width=\textwidth]{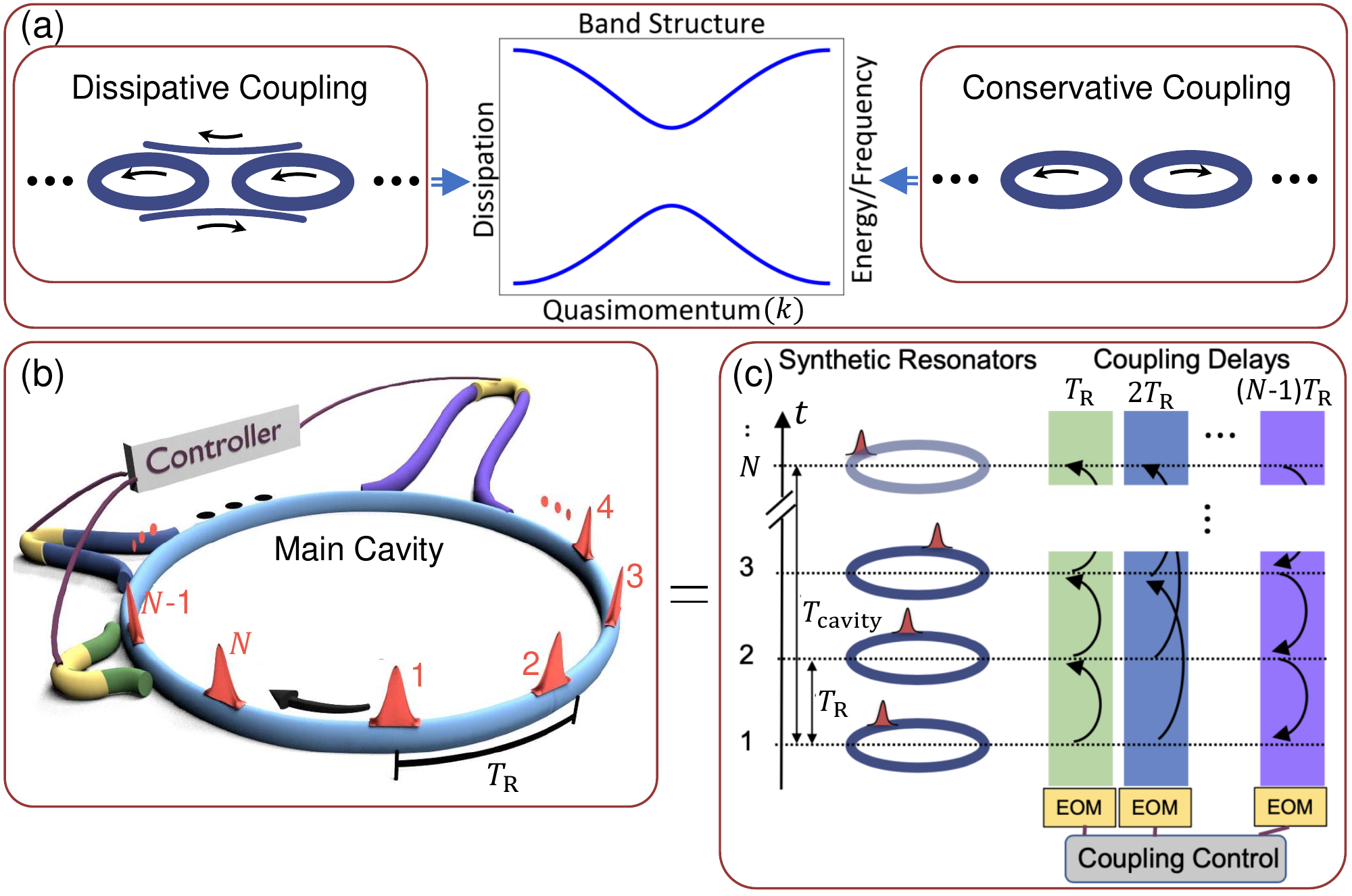}
    \caption{\textbf{Topological dissipation and time-multiplexed resonator networks}~\textbf{(a)} Dissipatively coupled resonators (left) can map the energy spectrum of a tight-binding Hamiltonian to the dissipation spectrum of the resonators. Conservatively coupled resonators (right) map the energy bands of a tight-binding Hamiltonian to the frequency spectrum of the resonators.~\textbf{(b)}~Schematic of a resonant cavity loop (light blue) that supports $N$ pulses separated by a repetition period, $T_{\rm R}$, and possesses delay lines of various lengths. The delay lines contain electro-optic modulators (EOMs, yellow), driven by a controller.
    \textbf{(c)} Equivalent synthetic resonator representation of (b). Each synthetic resonator consists of a single, recirculating pulse. The time-multiplexed network is built by coupling the pulses with delay lines, which are indicated by the shaded boxes.} 
    \label{fig:schematic_concept}
\end{figure*}

Most topological phases rely on conservative couplings to achieve nontrivial topological invariants. Conservative couplings arise when the elements of a system, be they the atoms of a quantum system or the ring resonators of a photonic system [Fig.~\ref{fig:schematic_concept}\textcolor{red}{(a)}], directly exchange information through their nonlocal, overlapping modes. As this direct coupling conserves energy, conservatively coupled systems are naturally described by energy spectra (in quantum mechanics) or frequency spectra (in photonics). In the topological lattice models studied in condensed matter physics~\cite{hasan_colloquium_2010}, cold atoms~\cite{cooper_cold_atoms_2019}, and photonics~\cite{ozawa_topological_2019}, these spectra form discrete bands in momentum space that are characterized by quantized, nonzero topological invariants. A central result of topological physics, the \textit{bulk-boundary correspondence}~\cite{jackiw_solitons_1976,hatsugai_chern_1993}, relates the presence of these nontrivial topological invariants to the existence of topologically protected edge states at the boundaries of the topological lattice. These edge states possess energies or frequencies that lie in the gap between two bands, and they are robust against disorder in the system as long as the disorder does not close the band gap. 

Dissipative couplings, on the other hand, indirectly couple the elements of a system through an intermediate reservoir~\cite{metelmann_2015,mukherjee_dissiaptive_2017}. The reservoir typically consists of a bath of environmental modes, and, in photonics, this may be the modes of a bus waveguide between two ring resonators~\cite{ding_2019}~[Fig.~\ref{fig:schematic_concept}\textcolor{red}{(a)}]. Dissipative couplings play a central role in superconducting circuits, ultracold atoms, and photonics, where they are used for reservoir engineering~\cite{barreiro_quantum_sim_2011}, laser mode-locking~\cite{haus_mode-locking_2000_w_url, wright_2020}, and quantum and photonic computing~\cite{verstraete_dissipation_2009, marandi_network_2014, inagaki_coherent_2016}. Several recent studies have proposed combining dissipative and conservative couplings to enable time-reversal symmetry breaking interactions~\cite{fang_generalized_2017} and to provide novel means to induce nontrivial topological invariants~\cite{wanjura_topological_2020, yoshida_diffusive_topological_2020, gneiting2020unraveling, dasbiswas_topological_2018, li_topological_2019}. These proposals suggest that dissipative coupling, like nonlinear~\cite{mukherjee_topological_solitons_2020,maczewsky_nonlinear_topology_2020, xia2020nonlinear} and non-Hermitian~\cite{bandres_topological_laser_2018,zhao_steering_2019, weidemann_funneling_2020, zhao_topological_2018} phenomena, may enable new topological phases and topology-inspired technologies for quantum and classical applications. However, to the best of our knowledge, the topological phases that arise due to purely dissipative couplings remain unrealized.

In this paper we introduce a highly scalable network  of time-multiplexed resonators to experimentally realize the dissipatively coupled analogs of two quintessential topological lattice models: the Su-Shrieffer-Heeger (SSH) model and the Harper-Hofstadter (HH) model. Our network's dissipative couplings engender a spectrum of dissipation bands that, like the energy bands of a conservatively coupled topological system, possess nontrivial topological invariants. Applied to our network, the bulk-boundary correspondence guarantees the existence of robust topological edge states with \textit{dissipation rates} that lie in the gaps of the network's dissipative band structure~[see Supplementary Information, Sec.~IV]. The isolation of these dissipation rates from the surrounding bulk bands presents new opportunities to engineer the properties of open quantum systems and of photonic devices.

Our time-multiplexed resonator network architecture [Fig.~\ref{fig:schematic_concept}\textcolor{red}{(b,c)}] uses the concept of synthetic dimensions~\cite{ozawa_topological_synthetic_2019, yuan_synthetic_2018} to generate dissipatively coupled lattices capable of hosting nontrivial topological invariants.
The network in Fig.~\ref{fig:schematic_concept}\textcolor{red}{(b)} contains $N$ resonant optical pulses separated by a repetition period $T_{\rm R}$. Each time-multiplexed pulse constitutes a synthetic resonator [Fig.~\ref{fig:schematic_concept}\textcolor{red}{(c)}], and we define each pulse to represent a single site in a synthetic lattice. As the pulses traverse the primary fiber loop (the ``main cavity''), a portion of each pulse enters the network's $(N-1)$ delay lines. These delay lines act analogously to the waveguide buses shown in Fig.~\ref{fig:schematic_concept}(a) by mediating unidirectional couplings between the network's time-multiplexed resonators. The lengths of the $(N-1)$ delay lines are chosen so that each pulse couples to the other $(N-1)$ pulses in the network, while the electro-optic modulators (EOMs) in the delay lines determine the strengths and phases of the couplings~\cite{marandi_network_2014}. By mapping the couplings of the network to a particular lattice type, the network represents the connections of a synthetic lattice. Significantly, by simply redefining this mapping~\textemdash~by adjusting the driving signals of the EOMs~\textemdash~the same network can represent an entirely new lattice type. In this work, we demonstrate our network's ability to readily implement multiple synthetic dimensions, tunable boundary conditions, reconfigurable inhomogeneous couplings, and time-reversal symmetry breaking gauge potentials. Simultaneously achieving these behaviors presents a significant challenge to existing platforms for synthetic dimensions~\cite{yuan_synthetic_2018,ozawa_topological_synthetic_2019,lustig_synthetic_dimensions_2019,dutt_synthetic_dimensions_2020}$^{,}$\footnote{While previous studies in photonic synthetic dimensions have utilized time-multiplexed schemes~\cite{regensburger_paritytime_2012, chalabi_synthetic_2019_PRL}, these traveling-wave architectures relied on conservatively coupled fiber loops and functioned similarly to real-space waveguide arrays. The flexible site-to-site couplings of our resonator-based design are instead akin to that of the original optical Ising machine~\cite{marandi_network_2014}}.

\begin{figure*}
   \centering
    \includegraphics[width=\textwidth]{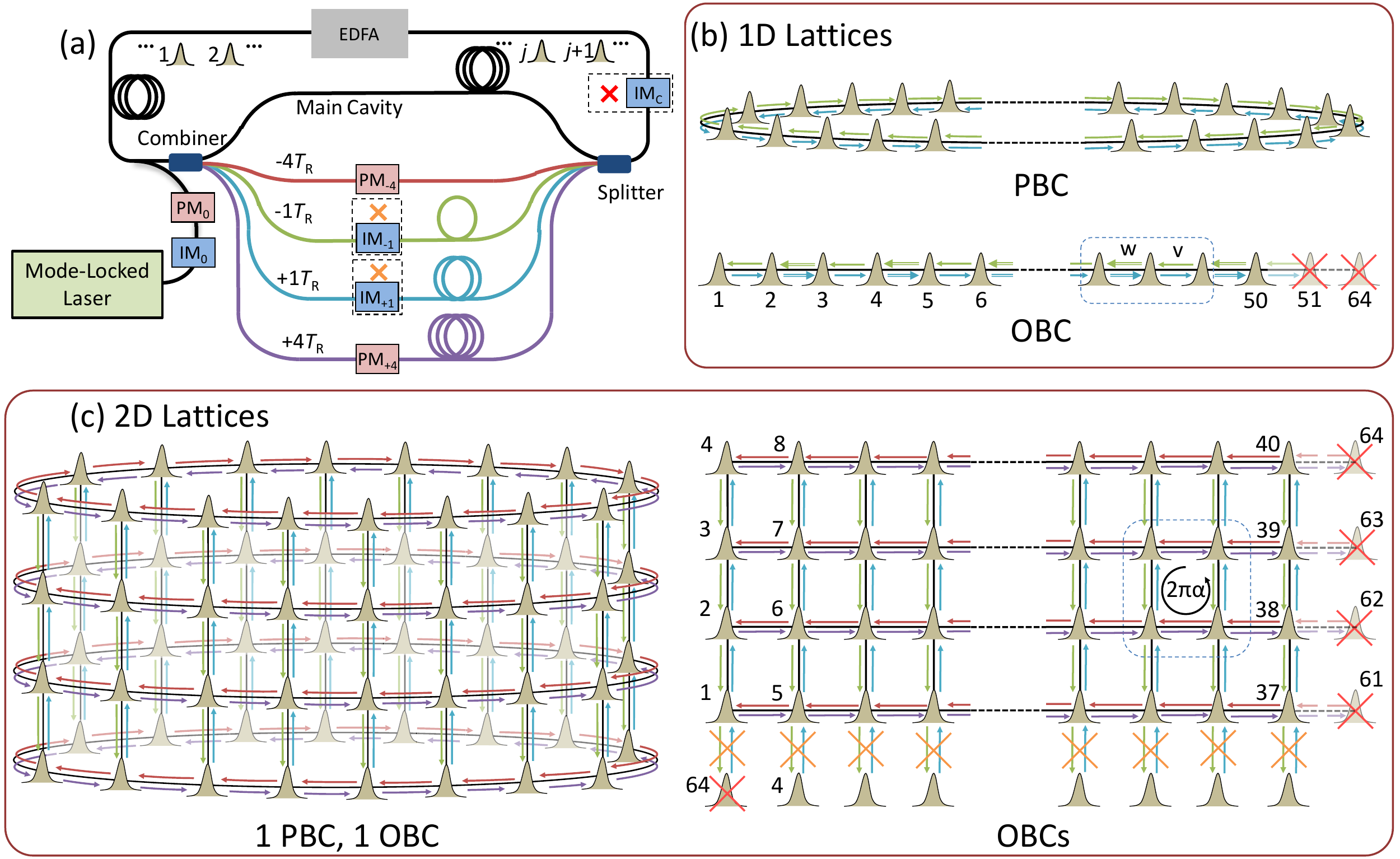}
    \caption{\textbf{Realizing 1D and 2D synthetic lattices with switchable boundary conditions in a time-mutliplexed resonator network}~\textbf{(a)}~We construct a four-delay line, time-multiplexed network capable of implementing two synthetic dimensions. \textbf{(b)}~With the $\pm4T_{\rm{R}}$ delay lines blocked, we use the intensity modulators (IM) in the $\pm{T_{\rm R}}$ delay lines, IM$_{\pm1}$, to implement a 1D chain with the staggered couplings ($w$ and $v$) of the SSH model. The intercavity IM, IM$_{\rm{C}}$, enables both periodic boundary conditions (PBCs) and open boundary conditions (OBCs). \textbf{(c)}~With all four delay lines, the network can implement a 2D square lattice. The PMs in the $\pm4T_{R}$ delay lines, PM$_{\pm4}$, produce the time-reversal symmetry breaking couplings of the Harper-Hofsatdter model, while IM$_{\pm1}$ enforces OBCs along the ``vertical'' direction. IM$_{\rm{C}}$ enables a PBC or an OBC along the ``horizontal'' direction.}
    \label{fig:lattice_mapping}
\end{figure*}

To study the dissipatively coupled equivalents of the SSH and HH models, we construct the four-delay-line network shown in Fig.~\ref{fig:lattice_mapping}\textcolor{red}{(a)}, which hosts $N=64$ synthetic lattice sites and can implement 1D chains and 2D square lattices with either open boundary conditions (OBCs) or a single periodic boundary condition (PBC)~[Figs.~\ref{fig:lattice_mapping}\textcolor{red}{(b,c)}]. We model the dynamics of this network by the general Lindblad master equation
\begin{equation}
    \frac{d}{dt} \rho = \mathcal{L}\rho = -i[\mathcal{H},\rho] + \sum_j \mathcal{D}[L_j] \rho,
    \label{eq:lindblad}
\end{equation}
\noindent where $\mathcal{H}$ denotes the Hermitian Hamiltonian dynamics due to conservative couplings between the sites labeled by $j$. Because our network possesses purely dissipative couplings, here $\mathcal{H}=0$. Instead, the dissipator $\mathcal{D}[L_j]\rho = L_j \rho L_j^\dagger - \{L_j^\dagger L_j, \rho\}/2$, with nonlocal jump operators $L_j$,
describes the purely dissipative couplings between the synthetic lattice sites~\cite{wanjura_topological_2020}. In Methods~Section~\ref{sec:master_eq}, we derive the jump operators for our network's delay line architecture~[Fig.~\ref{fig:schematic_concept}\textcolor{red}{(b)}] and show how to realize the inhomogeneous couplings and synthetic gauge potentials that enable us to construct our synthetic SSH and HH lattices. We also show how to engineer the phases of the delay lines to emulate purely conservative and hybrid conservative-dissipative dynamics with our network.

Starting from Eq.~\eqref{eq:lindblad}, we can express the evolution of the pulse amplitudes, $\langle a_{j}\rangle$, as 

\begin{equation}
    \frac{d\left<\textbf{a}\right>}{dt} = \left( K-\gamma \right)\left<\textbf{a}\right> + \textbf{P},
\label{eq:motion}
\end{equation}

\noindent where $t$ represents the slow-time (roundtrip-to-roundtrip) evolution of the network, $\gamma$ represents the intrinsic losses of the resonators, $\textbf{P}$ models a coherent drive, and $K$ is the network's coupling matrix. By engineering $K$ to implement the couplings of either the SSH or the HH model, our dissipatively coupled network acquires a dissipation spectrum identical to the topologically nontrivial band structure of the model under study. Therefore, in the presence of OBCs, our network is guaranteed to possess the same topological edge states as its conservatively coupled counterparts.

Properly selecting the coherent drive, $\textbf{P}$, allows us to probe specific states in our dissipative topological lattice. We generate the desired $\textbf{P}$ by using the modulators IM$_{\rm 0}$ and PM$_{\rm 0}$ in Fig.~\ref{fig:lattice_mapping}\textcolor{red}{(a)} to excite a specific state of the network, and we track the state's evolution to glean information about the underlying synthetic lattice. For instance, to probe the network's topological edge state, we program $\textbf{P}$ to excite the edge state. If the edge state is an eigenstate of the network, then the edge state excitation will remain localized as it resonates within the network. On the other hand, if the edge state is not an eigenstate, then the edge state excitation will undergo diffusive dynamics dictated by Eq.~\eqref{eq:motion}. Similarly, when we institute the network's single PBC, we can choose $\textbf{P}$ to excite a lattice's complete set of Bloch eigenstates. Measuring the steady states of these eigenstates allows us to reconstruct the 1D band structure of the model under study~[see Supplementary Information Sec.~II].

\begin{figure*}
    \centering
    \includegraphics[width=\textwidth]{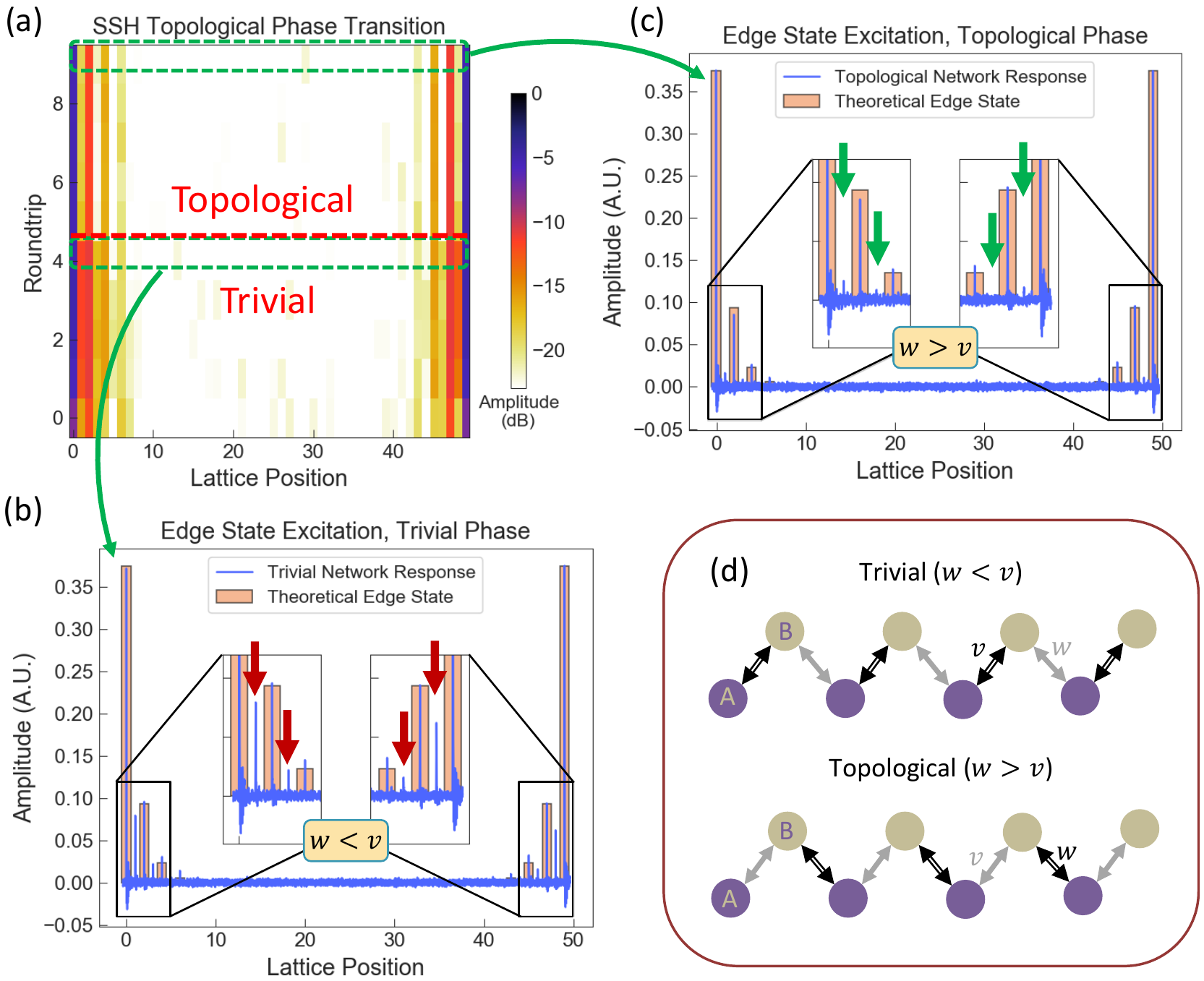}
    \caption{\textbf{Observations of the SSH edge state and a topological phase transition.}~\textbf{(a)}~For 10 roundtrips, we excite the SSH edge state corresponding to a coupling ratio of $w/v=\sqrt{2}$ in our time-multiplexed network while we tune the delay line couplings to induce a topological phase transition between the trivial and topological phases of the SSH model. \textbf{(b)} For the first 5 roundtrips, we set the network coupling ratio to the trivial phase ($w/v=1/\sqrt{2}$) and we observe that the edge state diffuses away from the edge as it resonates in the network (indicated by the thick red arrows). \textbf{(c)} After 5 roundtrips, we switch the coupling ratio to the topological phase ($w/v=\sqrt{2}$). The strong localization of the edge state in the topological phase (indicated by the thick green arrows) suggests that the edge state is an eigenstate of the network and corroborates our observation of a dynamic topological phase transition. \textbf{(d)}~Depictions of the SSH lattice corresponding to the trivial and topological phases.}

    \label{fig:ssh_edge}
\end{figure*}

To demonstrate purely dissipative topological phenomena, we first program our network to implement the couplings of the SSH model~\cite{su_solitons_1979, asboth_2015}. The SSH model describes a 1D dimerized chain with intra-dimer coupling $w$ and inter-dimer coupling $v$~[Fig.~\ref{fig:ssh_edge}], and the model's band structure is characterized by a $\mathbb{Z}_{2}$ topological invariant known as the winding number, $\mathcal{W}$. When $w<v$, $\mathcal{W}=0$, and the system is in a topologically trivial phase. However, when $w>v$, $\mathcal{W}=1$, and the system is in a topological phase that hosts mid-gap, topologically protected edge states.

We probe the SSH model's topological edge state by implementing a 50-site SSH lattice with OBCs and by inducing a dynamical topological phase transition between the SSH model's trivial and topological phases~[Fig.~\ref{fig:ssh_edge}\textcolor{red}{(a)}]. For 10 roundtrips, we excite the network with the SSH edge state corresponding to the coupling ratio $w/v=\sqrt{2}$. For the first 5 roundtrips, we prepare our SSH lattice in the trivial phase by setting the coupling ratio of the network to $w/v=1/\sqrt{2}$; for the remaining 5 roundtrips, we switch the synthetic lattice into the topological phase by changing the coupling ratio to $w/v=\sqrt{2}$. As shown in Fig.~\ref{fig:ssh_edge}\textcolor{red}{(b)}, when the lattice's couplings are in the trivial phase, the edge state excitation diffuses into the initially unoccupied states of the lattice. In contrast, when the couplings are in the topological phase, the edge state excitation remains strongly localized in the theoretically predicted edge state. This localization confirms the existence of a purely dissipative topological edge state in our time-multiplexed resonator network.

\begin{figure*}
    \centering
    \includegraphics[width=\textwidth]{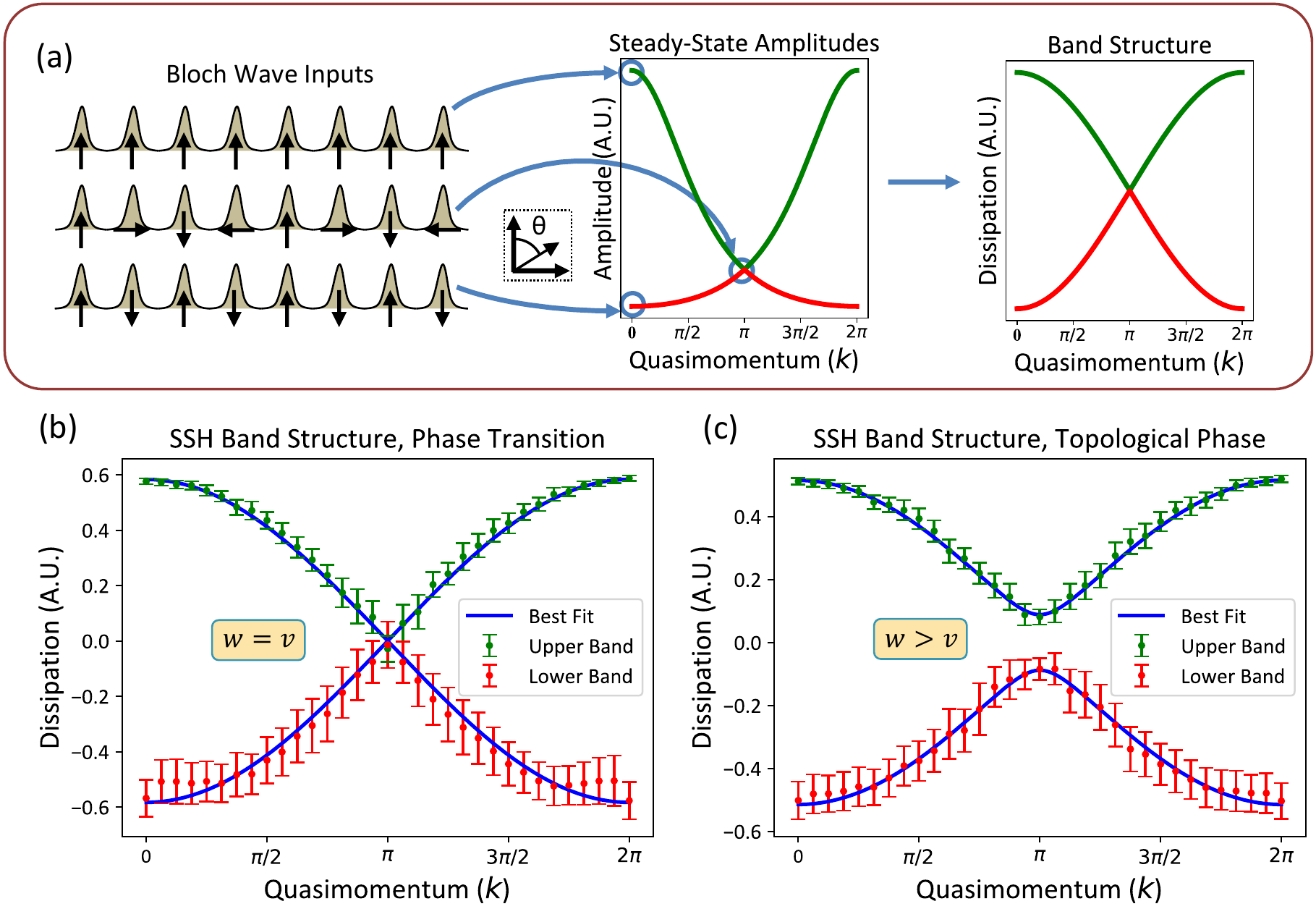}
    \caption{\textbf{Measurements of the SSH band structure.}~\textbf{(a)}~To measure the SSH band structure at a given coupling ratio, $w/v$, we excite the complete set of Bloch wave eigenstates in the network and measure the steady-state amplitude of each state. Using Eq.~\textcolor{red}{(}\ref{eq:motion}\textcolor{red}{)}, we transform these amplitudes into the SSH band structure. \textbf{(b)},\textbf{(c)}~Measurements of the SSH band structure at the phase transition point and in the topological phase. Using the best fit curves, we extract the coupling ratios for each measurement. We measure $w/v\approx1.0$ (expected $w/v=1$) and $w/v\approx1.4$ (expected $w/v=\sqrt{2}\approx1.414$) for the measurements in \textbf{(b)} and \textbf{(c)}, respectively. Note that the error bars represent the standard deviations of the measured dissipation rates.}

    \label{fig:ssh_bands}
\end{figure*}

Next, we experimentally reconstruct the dissipation bands of our SSH lattice in the topological phase ($w/v=\sqrt{2}$) and at the phase transition point ($w/v=1$). For both coupling ratios, we implement a 64-site SSH lattice with PBCs. In each case, we sequentially excite the network with each of its 64 Bloch eigenstates, and we measure the steady-state amplitude of each state. Then, using the fit procedure described in Supplementary Information Sec.~IIB, we extract the dissipation spectra from the measured steady state amplitudes [Figs.~\ref{fig:ssh_bands}\textcolor{red}{(b,c)}]. To evaluate the quality of our band structure measurements, we compare the coupling ratios, $w/v$, extracted from our fit procedure with the expected coupling ratios. This comparison provides a suitable metric for the quality of our measurements because, up to a constant, the SSH coupling ratio completely determines the band structure~\cite{asboth_2015}. For the band structure at the phase transition point (expected $w/v=1$), we measure $w/v\approx1.0$, while for the band structure in the topological phase (expected $w/v=\sqrt{2}\approx1.414$), we find $w/v\approx1.4$. Combined with our edge state measurements, the excellent agreement between the measured dissipation bands and the theoretical SSH band structure confirms that our network exhibits a topologically nontrivial dissipation spectrum.

\begin{figure*}
    \centering
    \includegraphics[width=\textwidth]{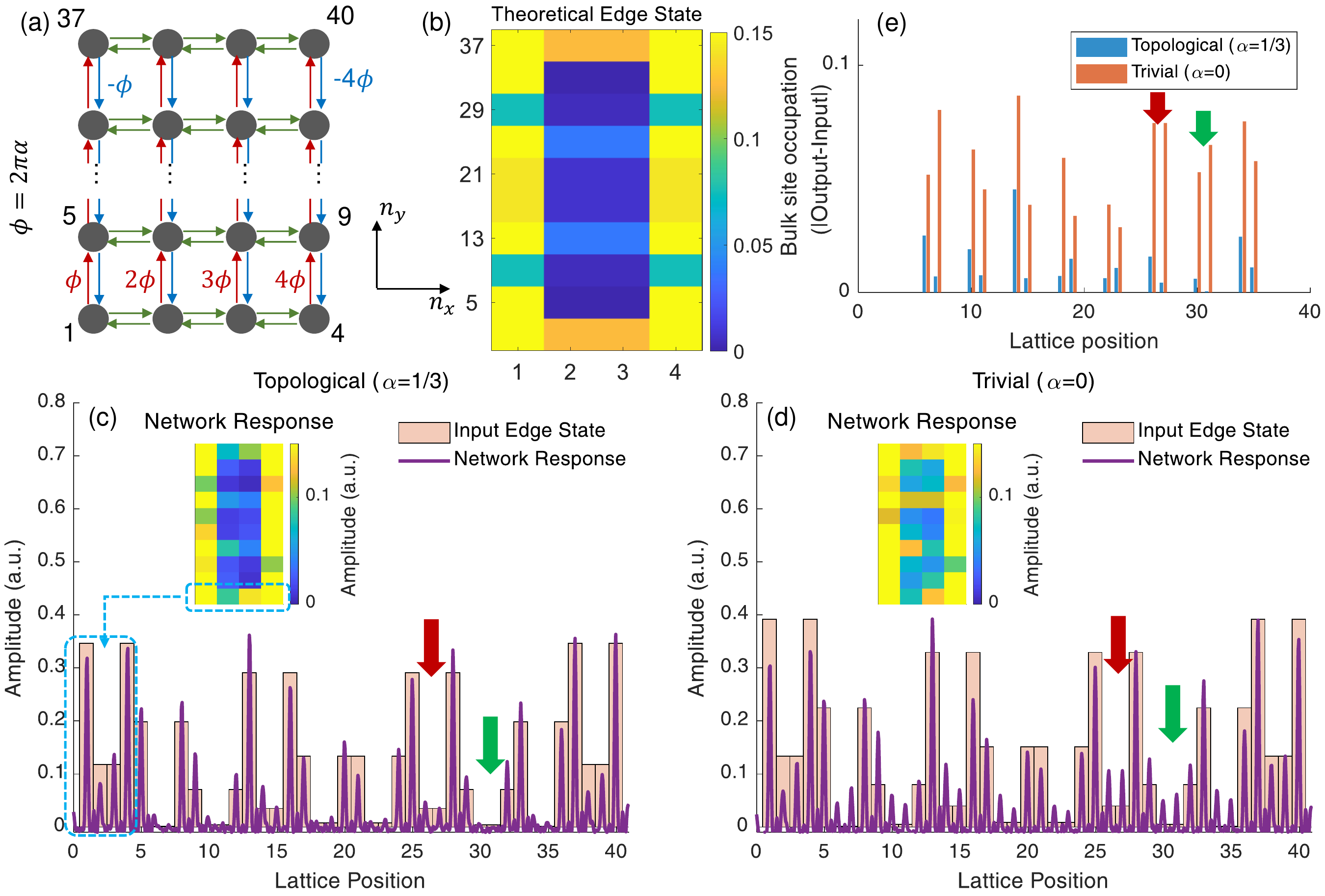}
    \caption{\textbf{Measurement of the Harper-Hofstadter edge state}. \textbf{(a)} Schematic of the Harper-Hofstadter (HH) model with a magnetic flux of $\phi = 2\pi\alpha$ per plaquette. The $n_x$-couplings do not have any phase, while the $n_y$-couplings implement the Peierl's phases $\phi, 2\phi, 3\phi, \cdots$ (red) and $-\phi, -2\phi, \cdots$ (blue).  {\bf (b)} One of the theoretical HH edge states of a 4$\times$10 lattice, which we inject into the network (also shown in (c) and (d) by light shaded rectangles). The plots in \textbf{(c)} and \textbf{(d)} contain a time trace of the measured steady-state pulse amplitudes of the network. As indicated by the arrows in \textbf{(c)}, each set of four pulses maps to one row in the inset color map, with earlier pulses in the time traces corresponding to lower rows in the color maps. \textbf{(c)}~When the phase modulators, PM$_{\pm4}$ implement the coupling phases of the HH model with $\alpha=1/3$, the edge state is an eigenstate of the network, and it resonates in the system. \textbf{(d)}~When we turn off these coupling phases to achieve $\alpha=0$, the edge state no longer resonates in the network. In particular, notice that light leaks into the ``bulk'' in the time trace of this measurement. {\bf(e)} The difference between the bulk site occupation for the topological case (blue) and the trivial case (orange). The thick red and green arrows in (c)-(e) indicate the bulk sites with the highest contrast between the topological and trivial cases.}
    \label{fig:hh_meas}
\end{figure*}

Finally, to showcase the scalability and flexibility of our time-multiplexed network architecture, we reconfigure our synthetic lattice to probe the topological edge state of the HH model. The HH model describes a 2D square lattice subjected to a perpendicular magnetic field~\cite{ozawa_topological_2019, hofstadter_1976,harper_1955}, whose strength is characterized by a dimensionless magnetic field parameter, $\alpha$. For rational $\alpha$, the bands of the HH model acquire a nonzero topological invariant known as the Chern number, $\mathcal{C}$, which gives rise to topologically protected edge states~\cite{hatsugai_chern_1993}. In our network we use the modulators PM$_{\pm4}$ in Fig.~\ref{fig:lattice_mapping}\textcolor{red}{(a)} to achieve an effective magnetic field with $\alpha=1/3$ in a $4\times10$ synthetic HH lattice with OBCs~[Fig.~\ref{fig:hh_meas}\textcolor{red}{(a)}]. Because the dissipative couplings of our network are unidirectional, the synthetic magnetic field generated by the delay lines breaks time-reversal symmetry~\textemdash~\textit{meaning that our network possess truly nonzero Chern numbers}. This is in stark contrast with earlier optical implementations of the HH model, which either do not break time-reversal symmetry~\cite{hafezi_imaging_2013} or only break $z$-reversal symmetry~\cite{rechtsman_2013}. As shown in Fig.~\ref{fig:hh_meas}\textcolor{red}{(c,e)}, when we excite topologically protected edge state of the HH model in the presence of the synthetic magnetic field, the edge state remains well localized. However, if we excite the edge state in the absence of the synthetic magnetic field ($\alpha=0$), the lattice represents a trivial insulator, and the edge state diffuses into the ``bulk'' of the synthetic lattice. We quantify the contrast between the trivial and topological phases by defining a bulk occupation fraction $f_{\, \rm bulk} = \sum_{n_x,n_y\,  \in\, \rm\, bulk} |\psi_{n_x,n_y}|^4 $, subject to the normalization $\sum_{n_x, n_y} |\psi_{n_x, n_y}|^2 = 1$. We calculate $f^{\rm topo}_{\rm bulk} = 5.6\times10^{-4}$ and $f^{\rm trivial}_{\rm bulk}=2.2\times10^{-3}$ for the topological and trivial phases respectively. As our $4\times10$ lattice has 16 bulk sites and 24 edge sites, $f^{\rm trivial}_{\rm bulk}/f^{\rm topo}_{\rm bulk} \approx 4$ indicates a significantly stronger penetration into the bulk for the lattice in the trivial phase. Based on this observation, we conclude that our time-multiplexed synthetic HH lattice hosts a multidimensional topological edge state.

Our dissipatively coupled implementations of the 1D SSH model and the 2D HH model represent the first experimental realizations of topological phenomena in the presence of purely dissipative couplings. We leverage our time-multiplexed network's dissipative dynamics for edge state and band structure measurements, and we utilize the time-reversal symmetry breaking nature of our dissipative couplings to introduce nonzero Chern numbers. Our time-multiplexed resonator architecture also offers a promising platform for future work in synthetic dimensions. Our design can be extended to lattices in higher than two dimensions~\cite{zhang_four-dimensional_2001_corrected_url, petrides_six-dimensional_2018, lohse_exploring_2018, zilberberg_photonic_2018, wang_multidimensional_synthetic_2020} and to lattices with long-range couplings~\cite{bell_spectral_2017}, can achieve dense connectivity between lattice sites, and can realize dynamic and inhomogeneous synthetic gauge fields~\cite{fang_controlling_2013}~\textemdash~a combination that is not easily achievable with other experimental platforms. We anticipate that dissipative couplings will enable new topological devices with applications to quantum computing and photonics. Immediate extension of our current experiments include exploring non-Hermitian and nonlinear topological behaviors in dissipatively coupled time-multiplexed networks.

\section*{Acknowledgements}

The authors are grateful to Michael Fraser of NTT Research for insightful discussions. The authors acknowledge support from ARO Grant No. W911NF-18-1-0285 and NSF Grants No. 1846273 and 1918549. S.F. acknowledges the support of a Vannevar Bush Faculty Fellowship from the U.S. Department of Defence (Grant No. N00014-17-1-3030).  L.Y. acknowledges the support of the National Natural Science Foundation of China (11974245). F.N. acknowledges support from ARO (W911NF-18-1-0358), JST-CREST (JPMJCR1676), JSPS (JP20H00134), AOARD (FA2386-20-1-4069), and FQXi (FQXi-IAF19-06). The authors wish to thank NTT Research for their financial and technical support.

\pagebreak

\bibliography{bibliography.bib, zotero-refs-ad_manual.bib, manual_ad_refs.bib}

\section*{Methods}

\subsection{Network Architecture}\label{sec:architecture}

The time-multiplexed optical network studied in this work hosts $N=64$ time-multiplexed resonators and possesses four delay lines, labeled the $\pm{T_{\rm{R}}}$ and the $\pm4T_{\rm{R}}$ delay lines~[Fig.~\ref{fig:lattice_mapping}\textcolor{red}{(a)}]. Each delay line differs in length from the corresponding section in the main cavity by an integer multiple of the pulse repetition period $T_{\rm{R}}$. The ``$-$'' (``$+$") indicates that the delay line is shorter (longer) than the corresponding main loop section, and the accompanying number denotes the range of the coupling (e.g., the $\pm4T_{\rm{R}}$ delay lines implement fourth-nearest neighbor coupling). As the separate $\pm{NT_{\rm{R}}}$ delay lines provide independent control over each direction of the $N$th-nearest neighbor couplings, it is straightforward to implement nonreciprocal couplings between sites.

The Su-Schrieffer-Heeger (SSH) model only requires nearest-neighbor coupling, so to study the SSH model, we block the $\pm4T_{\rm{R}}$ delay lines. We then map the pulses in the main cavity to the one-dimensional chain pictured in Fig.~\ref{fig:lattice_mapping}\textcolor{red}{(c)}, where the colors of the couplings correspond to the colors of the delay lines that implement them. The intensity modulators inserted in the $\pm{T_{R}}$ delay lines provide pulse-to-pulse control over the coupling strengths of each delay line and enable us to implement the staggered couplings of the SSH model. Moreover, while the topology of the main cavity lends itself to periodic boundary conditions (PBCs), the intensity modulator (IM) within the main cavity (IM$_{\rm C}$ in Fig.~\ref{fig:lattice_mapping}\textcolor{red}{(a)}) provides control over the boundaries of the synthetic 1D lattice. We can switch from a PBC to an open boundary conditions (OBC) simply by using IM$_{\rm{C}}$ to suppress time-slots in the main cavity~[Fig.~\ref{fig:lattice_mapping}\textcolor{red}{(c)}].

To realize the HH model in the network of Fig.~\ref{fig:lattice_mapping}\textcolor{red}{(a)}, we use the $\pm4T_{\rm{R}}$ delay lines to define nearest-neighbor couplings along the second dimension of a synthetic square lattice~\cite{yuan_synthetic_space_2018}. By using the IMs in the $\pm{T_{R}}$ delay lines to suppress the ``spiraling'' boundary condition along this second synthetic dimension, we arrive at the lattices presented in Fig.~\ref{fig:lattice_mapping}\textcolor{red}{(c)}, where, once again, IM$_{\rm{C}}$ enables us to implement either a strip with a single PBC or a square lattice with OBCs. To achieve the time-reversal symmetry breaking coupling phases of the HH model, we place phase modulators (PMs) in the $\pm4T_{\rm{R}}$ delay lines. We utilize our independent control over each delay line to introduce a synthetic magnetic flux in each plaquette of the synthetic lattice~[Fig.~\ref{fig:lattice_mapping}\textcolor{red}{(c)}].

For the measurements presented in the main text, we probe the properties of the network by exciting states in the network and recording the network's steady-state response. To excite the desired edge states and Bloch wave eigenstates, we use an IM (IM$_{0}$) and a PM (PM$_{0}$) at the input to the main cavity~[Fig.~\ref{fig:lattice_mapping}\textcolor{red}{(a)}]. These modulators encode the intensities and phases of the desired state onto a train of pulses from a mode-locked laser. Upon entering the cavity, these pulses excite the sites of the synthetic lattice (i.e. the time bins of the network) with particular amplitudes and phases. By exciting each site repeatedly over multiple roundtrips of the network, we bring the cavity to a resonant steady-state condition.

\subsection{Derivation of the Master Equation}\label{sec:master_eq}

In this section, we derive the Lindblad master equation for the dynamics of $N$ time-multiplexed resonators with optical delay line couplings. For simplicity, we will consider the case of a network with $\pm T_R$ delay lines, which can represent 1D models with nearest-neighbor coupling.

The finite difference equations for the $j$th time-multiplexed resonator $a_j$ can be written as,
\begin{equation}
\begin{split}
    a_j(t+T_{\rm{rt}}) = (1-t_1^2)\, (1-t_2^2)\, a_j(t) + (it_1)^2 e^{i\phi_1}\, a_{j+1}(t) \\ + (it_2)^2 e^{i\phi_2} \, a_{j-1} (t),
\end{split}
\end{equation}

\noindent where $T_{\rm{rt}}=NT_{\rm R}$ is the duration of one roundtrip through the main cavity, $t_{1}$ ($t_{2}$) and $\phi_{1}$ ($\phi_{2}$) represent the coupling strength and coupling phase of the the $-T_{\rm R}$ ($+T_{\rm R}$) delay line, respectively.

Assuming that the change in the field over the course of roundtrip is small, which is the case for $t_1^2, t_2^2 \ll 1$, we obtain,
\begin{equation}
\begin{split}
    T_{\rm{rt}}\,  \dot a_j = (-t_1^2 - t_2^2)\, a_j(t) - t_1^2 e^{i\phi_1} a_{j+1}(t) - t_2^2 e^{i\phi_2} a_{j-1}(t) \\ + {O}(t_1^2 t_2^2).
\end{split}
\end{equation}

From this equation, we can read off the effective inter-site coupling terms as

\begin{subequations}
\begin{equation}
K_{j,j+1} = -it_1^2 e^{i\phi_1}/T_{\rm rt},
\end{equation}
\begin{equation}
K_{j+1,j} = -it_2^2 e^{i\phi_2}/T_{\rm{rt}}.
\end{equation}
\label{eq:inter_site_coupling}
\end{subequations}

Following the treatment of Ref.~\onlinecite{wanjura_topological_2020}, we can express the (Hermitian/conservative) Hamiltonian resulting from Eqs.~\eqref{eq:inter_site_coupling} as $\mathcal{H} = \sum_j (J a^\dagger_j a_{j+1} + J^* a_j a^\dagger_{j+1})$, where

\begin{equation}
    J = (K_{j,j+1} + K^*_{j+1,j})/2.
\end{equation}

Note that $J$ is in general complex, since we have not made any assumptions about $t_1, t_2, \phi_1,$  and $\phi_2$.

Similarly, we can write the anti-Hermitian, or dissipative coupling, between sites as as,
\begin{equation}
    \Gamma e^{-i\theta} = (K_{j,j+1} - K^*_{j+1,j}).
\end{equation}

The classical analog of the Lindblad master equation for the system density matrix in the Sch\"odinger picture for such a time-multiplexed resonator network is then,
\begin{equation}
    d\rho /dt  = -i[\mathcal{H}, \rho] + \sum_j \mathcal{D}[L_j] \rho,
\end{equation}
where the dissipator $\mathcal{D}[L_j]\rho = L_j \rho L_j^\dagger - \{L_j^\dagger L_j, \rho\}/2$ and the nonlocal jump operator $L_j = \sqrt{\Gamma}\, (a_j + e^{-i\theta}a_{j+1})$ has a dissipative coupling rate $\Gamma$ between neighboring sites.

For the simplest case of a 1D lattice with $t_{1}=t_{2}=t$, the nonlocal jump operator $L_j = \sqrt{\Gamma}\, (a_j + e^{-i\theta}a_{j+1})$ acquires a dissipative coupling rate of $\Gamma = (2t_1^2/NT_R) \cdot \cos ((\phi_1-\phi_2)/2)$. Here $\theta = (\pi-\phi_1-\phi_2)/2$ plays the role of a gauge potential~\cite{fang_realizing_2012} that depends on the phase $(\phi_1+\phi_2)$ introduced by the delay lines. In the 2D case, by properly preparing the distribution of phases, we can construct the effective magnetic field perpendicular to the 2D synthetic lattice. Furthermore, the conservative Hamiltonian coupling strength is $|J| = (t_1^2/NT_R)\sin((\phi_1-\phi_2)/2)$, and hence by setting $\phi_1=\phi_2$, we realize purely dissipative Lindbladian dynamics ($\mathcal{H} = 0$) [see Supplementary Information Sec.~V for details].

\subsection{Measurement Procedures}\label{sec:measurement}

\subsubsection{SSH Band Structure Measurements}\label{sec:band_meas}

To measure the SSH band structure, we generate the modulator driving signals to implement (1) the desired coupling ratio within the network (using IM$_{\pm1}$), and (2) the Bloch wave excitations at the input to the cavity (using PM$_{0}$). By not using the intercavity IM, IM$_{\rm{C}}$, the network inherently implements a PBC, so that we implement a 64 pulse (32 dimer) SSH lattice.

To execute the experiment, we excite each Bloch eigenstate in the network and record the network's steady-state response to each Bloch state. We repeat this measurement 5 times for each Bloch wave and compile the data from the different measurements to generate a plot of the mean steady-state amplitudes versus wavevector. We then solve Eq.~\textcolor{red}{(}\ref{eq:motion}\textcolor{red}{)} to relate the steady state amplitudes of the Bloch waves, $|c(k)|^{2}$, to the dissipation eigenvalues of the SSH model. We find

\begin{equation}
    |c(k)|^{2} = \frac{A}{\left(\gamma-\lambda_{\text{SSH}}\right)^{2}}+d,
    \label{eq:spec_fit_model}
\end{equation}

\noindent where $\gamma$ is the network loss,

\begin{equation}
\lambda_{\text{SSH}}=\pm\sqrt{w^{2}+v^{2}+2wv\cos{\left(k\right)}},
\end{equation}

\noindent and $A$ and $d$ account for the detector scaling and bias, respectively.

We fit the measured amplitudes with a rescaled version of Eq.~\eqref{eq:spec_fit_model} using Markov chain Monte Carlo simulations~\cite{aster_2013}. We use the fit parameters to transform the measured amplitudes into the SSH band structures shown in Figs.~\ref{fig:ssh_bands}\textcolor{red}{(a),(b)}.

\subsubsection{Edge State Measurements}\label{sec:edge_meas}

Our edge state measurements follow a procedure similar to that of our band structure measurements. To observe the HH model's edge state, we first generate the modulator driving signals to implement the synthetic gauge field of the HH model (using PM$_{\pm4}$) and to produce the HH edge state at the input to the network (using IM$_{0}$ and PM$_{0}$). As suggested in Fig.~\ref{fig:lattice_mapping}\textcolor{red}{(c)}, we use the delay line IMs, IM$_{\pm1}$, to create OBCs along the one direction of the lattice, while IM$_{\rm{C}}$ produces OBCs along the other direction. The result is that we implement a finite, $4\times 10$ HH lattice with an effective synthetic magnetic field strength corresponding to $\alpha=1/3$.

To probe the topologically nontrivial state of our lattice, we excite the HH edge state in the network and record the system's steady-state. In the presence of the synthetic gauge field, the edge state is an eigenstate of the network, so the excited edge state resonates unperturbed within the network. This result is shown in Fig.~\ref{fig:hh_meas}\textcolor{red}{(c)}.

To confirm that the lattice hosts a 2D topological edge state, we next turn off the synthetic gauge field by turning off the driving signals on PM$_{\pm4}$. The network then implements a trivial $4\times 10$ square lattice. We excite the same topological edge state in the trivial lattice and observe that the network's steady state response deviates from the edge state excitation~[Fig.~\ref{fig:hh_meas}\textcolor{red}{(d)}]. This confirms that, in the presence of the synthetic gauge field, the topological edge state is an eigenstate of the network.

For the SSH model, we observe the topological edge state in the context of a dynamical topological phase transition between the trivial and topological phases~[Fig.~\ref{fig:ssh_edge}\textcolor{red}{(c)}]. We begin by generating the modulator driving signals to implement the SSH model's couplings and to excite the SSH edge state. In addition, we now use IM$_{\rm C}$ to implement a 50 pulse SSH lattice with OBCs. We excite the SSH edge state in the network for 10 roundtrips. For the first 5 roundtrips, we program the couplings so that the network is in the trivial phase of the SSH model. In this case, we observe that the steady-state response of the network deviates from the excited edge state. For the final 5 roundtrips, we switch the coupling strengths so that the network is in the SSH model's topological phase. Now we observe that the network response remains strongly localized in the edge state. This indicates that the topological edge state is an eigenstate of the network when the network is in the topological phase.

\end{document}